\documentclass[showpacs,nofootinbib,preprintnumbers,prd,superscriptaddress,twocolumn]{revtex4}

\usepackage{amsmath}
\usepackage{amsfonts}
\usepackage{amssymb}
\usepackage{graphicx}
\usepackage[titletoc]{appendix}
\usepackage{color}
\usepackage{hyperref}
\usepackage{cleveref}
\usepackage[rightcaption]{sidecap}
\usepackage{subfigure}
\usepackage{comment}
\usepackage{leftindex}
\usepackage{csquotes}
\usepackage{adjustbox}

\usepackage{mathtools}

\usepackage{dcolumn}
\usepackage{braket}
\usepackage{array}
\usepackage{ctable}
\usepackage{multirow}
\usepackage{siunitx}
\usepackage{longtable}
\usepackage{tabularx}
\usepackage{booktabs}

\graphicspath{{Graphics/}}

\def\be{\begin{equation}}
\def\ee{\end{equation}}
\def\bea{\begin{eqnarray}}
\def\eea{\end{eqnarray}}
\renewcommand{\d}{\mathrm{d}}

\definecolor{vividviolet}{rgb}{0.62, 0.0, 1.0}
\definecolor{amaranth}{rgb}{0.9, 0.17, 0.31}
\definecolor{palatinateblue}{rgb}{0.15, 0.23, 0.89}
\definecolor{brightpink}{rgb}{1.0, 0.0, 0.5}
\definecolor{cornflowerblue}{rgb}{0.39, 0.58, 0.93}
\definecolor{deepcarminepink}{rgb}{0.94, 0.19, 0.22}
\definecolor{radicalred}{rgb}{1.0, 0.21, 0.37}

\hypersetup{ linktoc=all,
    colorlinks, linkcolor={palatinateblue},
    citecolor={brightpink}, urlcolor={amaranth}
}

\begin{document}

\title{Gravitational dark matter production  from fermionic spectator fields during inflation}

\author{Alessio Belfiglio}
\email{alessio.belfiglio@unisi.it}
\affiliation{DSFTA, University of Siena, Via Roma 56, 53100 Siena, Italy.}

\author{Orlando Luongo}
\email{orlando.luongo@unicam.it}
\affiliation{School of Science and Technology, University of Camerino, Via Madonna delle Carceri, Camerino, 62032, Italy.}
\affiliation{SUNY Polytechnic Institute, 13502 Utica, New York, USA.}
\affiliation{Istituto Nazionale di Fisica Nucleare (INFN), Sezione di Perugia, Perugia, 06123, Italy.}
\affiliation{INAF - Osservatorio Astronomico di Brera, Milano, Italy.}
\affiliation{Al-Farabi Kazakh National University, Al-Farabi av. 71, 050040 Almaty, Kazakhstan.}

\begin{abstract}
    We investigate the gravitational particle production from vacuum for a minimally coupled fermionic spectator field during a single-field inflationary phase. We observe that metric perturbations arising from the quantum fluctuations of a scalar inflaton field enhance gravitational production, showing that such a perturbative contribution becomes dominant if the field mass is sufficiently smaller than the inflationary Hubble rate. We focus on modes that leave the Hubble horizon during the latest stages of slow-roll and we numerically compute the total number of particles obtained from perturbations, providing a lower bound on the amount of such \enquote{geometric} particles for the case of Starobinsky inflation and a quadratic hilltop scenario. Our outcomes are compatible with the net observational cold dark matter abundance as experimentally measured, whose dark matter candidate exhibits mass in the range $10^5 \lesssim m \lesssim 10^7$ GeV, excluded by previous non-perturbative calculations. 
\end{abstract}

\pacs{04.62.+v, 98.80.-k, 98.80.Cq, 03.67.bg}

\maketitle
\tableofcontents


\section{Introduction}\label{intro}

Dark matter (DM) nature remains elusive as no conclusive experiments have been able to definitively identify its properties yet \cite{RevModPhys.90.045002}. 

Recent attempts to detect weakly interacting massive particles (WIMPs) with masses ranging from a few to 100 GeV have unfortunately failed to positively yield results \cite{Bertone:2010at,PhysRevLett.118.021303,PhysRevLett.119.181302,PhysRevLett.121.111302}, shifting the interest to ultralight DM candidates, including axions  \cite{PhysRevD.78.083507,Sikivie:2009fv,PhysRevD.81.123530,PhysRevD.82.123508,RevModPhys.82.557,Marsh:2015xka} and axion-like particles, \cite{Graham:2015ouw,Niemeyer:2019aqm,Choi:2020rgn} or to superheavy ones such as Planckian interacting DM \cite{PhysRevLett.116.101302,Hashiba:2018iff,Garny:2017kha}, WIMPzillas \cite{Kolb:1998ki,Kolb:2007vd} and so on.

The inconclusive explanation of DM's nature is partly due to the unknown mechanism by which it emerges. Nevertheless, DM's existence mainly comes from gravitational effects on ordinary matter and radiation. Thus, it seems quite reasonable to consider whether \emph{gravity alone could have played a role in the creation of dark matter particles}.

Within this picture, a promising mechanism for DM generation is represented by  \emph{gravitational particle production} (GPP), generating particles directly from vacuum fluctuations, not requiring an interaction between the DM candidate and other quantum fields \cite{PhysRevLett.21.562,PhysRev.183.1057,PhysRevD.3.346,RevModPhys.96.045005}, and exhibiting reasonable results in various cosmological scenarios, such as inflationary stages \cite{PhysRevD.35.2955,PhysRevD.59.023501,PhysRevD.64.043503,PhysRevD.101.083516} and reheating \cite{PhysRevD.99.043008,Cembranos:2019qlm,PhysRevD.101.063529} phases.

Initial studies on GPP were performed by assuming an unperturbed background expansion, typically modeled by a Friedmann-Robertson-Walker spacetime, and then computing the field modes in the asymptotic past and future, related by appropriate Bogoliubov transformations \cite{Birrell_Davies_1982,Parker:2009uva}. In particular, gravitational fermion production has been recently studied by picturing the early universe evolution via an instantaneous transition from a de Sitter inflationary stage\footnote{The assumption of instantaneous transition after inflation may lead to an overestimate of GPP, see e.g. Ref. \cite{Li:2019ves}.} to the radiation dominated era \cite{Chung:2011ck,PhysRevD.101.123522}, showing that plausible DM candidates have mass $m \gtrsim 10^8$ GeV. 

However, the additional presence of inhomogeneities is able to \emph{enhance} the GPP mechanism from vacuum \cite{PhysRevD.39.389,PhysRevD.45.4428}, thus leading to possibly larger number densities. In this respect, a non-negligible perturbative GPP contribution was recently derived for a spectator scalar field in inflation, by coupling its energy-momentum tensor to the metric perturbations generated by a scalar inflaton field \cite{PhysRevD.110.023541}.

More generally, inflaton fluctuations generate perturbations within the inflationary quasi-de Sitter dynamics, and the presence of such inhomogeneities leads to the gravitational production of additional \emph{geometric} particles, which may also absorb momentum from the background gravitational field, thus introducing mode-mixing in GPP processes.

Given the importance of primordial perturbations for structure formation in the universe \cite{Mukhanov:1990me,Brandenberger:2003vk}, such a contribution cannot be neglected \emph{a priori} and hints toward a geometric DM origin may emerge, especially when quantum fields are coupled with spacetime curvature.
This has been considered in view of addressing the cosmological constant problem \cite{Belfiglio:2022qai} and extending the nature of DM constituents \cite{Belfiglio:2023rxb,Belfiglio:2024swy}, while possible signatures of mode-mixing may arise from the entanglement entropy associated with GPP processes \cite{PhysRevD.105.123523,PhysRevD.107.103512,PhysRevD.109.123520}.

Under specific circumstances, perturbative GPP from inhomogeneities may represent the dominant mechanism for gravitational production, as in the case of a minimally coupled spin-$1/2$ field with sufficiently small mass\footnote{A similar scenario arises for a conformally coupled scalar field; see e.g. Ref. \cite{RevModPhys.96.045005} on how GPP is related to conformal symmetry breaking.}. The GPP associated with a Dirac spectator field was investigated some time ago in the context of reheating \cite{Bassett:2001jg}, showing that perturbative effects are expected to dominate for field masses up to $m \simeq 10^7$ GeV, if metric perturbations are sufficiently amplified during the preheating stage. However, preheating and reheating stages are highly model-dependent, and a sufficient amplification of metric perturbations may take place only if the inflaton field is appropriately coupled to the particle standard model fields\footnote{For example, such amplification would not be possible in some recently proposed gravitational reheating scenarios \cite{PhysRevD.105.075005,PhysRevD.107.043531,Dorsch:2024nan}, where the alternative mechanism of gravity-mediated inflaton decay has been proposed to reheat the universe \cite{Ema:2015dka,Tang:2017hvq}.}.

Motivated by the above scenarios, we here aim to quantify the effects of scalar inflaton fluctuations on the GPP associated with a Dirac spectator  field during inflationary slow-roll. We model the inflationary epoch as a quasi-de Sitter stage, where a small and constant slow-roll parameter is introduced by fixing its value at the horizon crossing of a standard pivot scale $k_*=0.05$ Mpc$^{-1}$. The energy-momentum tensor of the fermionic field is then coupled to the metric perturbations generated by the scalar inflaton, excluding any possible non-gravitational coupling between the two fields. If the Dirac field mass is much smaller than the inflationary Hubble rate, the corresponding GPP number density can be evaluated via the Weyl tensor associated with inflationary perturbations \cite{PhysRevD.45.4428,Bassett:2001jg}. Hence, we show that, for sufficiently high inflationary energy scales, tuned on Planck's constraints \cite{Planck:2018jri}, perturbative GPP from vacuum may lead to a \emph{non-negligible number densities of fermionic particles}, and it may also account by itself for \emph{a cold DM candidate for a field mass range, $10^5 \lesssim m \lesssim 10^7$} GeV, i.e., where non-perturbative GPP from expansion is typically inefficient, since conformal symmetry breaking associated with the fermionic field mass is negligible.

In particular, we specify our calculation to two main models of inflation, passing the Planck's constraints. The first is the large-field Starobinsky scenario \cite{STAROBINSKY198099,PhysRevD.110.063552}, whereas the second is the small-field quadratic hilltop scenario \cite{Boubekeur:2005zm}. Within those paradigms, we demonstrate that dark matter may emerge, regardless the particular minimally-coupled field, involved into our computation. Accordingly, properties of such DM particles are also reported and discussed. More precisely, we focus on field modes that leave the Hubble horizon during the latest stages of slow-roll, highlighting how the number density spectrum of fermionic particles is strongly blue-tilted.  This result is typically in agreement with the non-perturbative GPP spectrum of minimally coupled fermions \cite{RevModPhys.96.045005} and may thus allow to avoid an unacceptably large value of DM isocurvature on CMB scales \cite{PhysRevD.91.043516}, which is constrained by observations of the CMB anisotropies \cite{Planck:2018jri}. 

Within the above-presented picture, we also discuss possible contributions arising from the reheating phase, which appear essential to consistently describe the overall process and to obtain a more precise estimate of the particle densities from purely gravitational effects. In particular, we remark that perturbative GPP from vacuum is not expected to provide a relevant contribution during the inflaton oscillations, while the inclusion of gravity-mediated inflaton decay may result in additional DM particles, despite this abundance heavily relies on the reheating details.


The paper is thus structured as follows.  In Sec. \ref{sec2}, we outline the inflationary scenario and the spectator field dynamics, also introducing the GPP mechanism from vacuum. In Sec. \ref{sec3}, we quantify perturbative GPP associated with the spectator fermionic field within both large and small-field inflation, highlighting how it represents a plausible DM candidate. Finally, in Sec. \ref{sec4}, we discuss the physical consequences of our findings and we draw our conclusions.


\section{Inflationary set up} \label{sec2}

Within the context of single-field inflation, we consider a scalsr field $\phi$, with Lagrangian density
\be \label{infl_lag}
\mathcal{L}_I= \frac{1}{2}  g^{\mu \nu} \phi_{, \mu} \phi_{,\nu}- V(\phi),
\ee
where the potential $V(\phi)$ drives the inflationary phase, while $g_{\mu \nu}$ denotes the metric tensor. The dynamics of the inflaton field is mostly studied via the standard ansatz \cite{Brandenberger:2003vk,Riotto:2002yw}
\be \label{infans}
\phi({\bf x},\tau)=\phi_B(\tau)+ \delta \phi ({\bf x},\tau),
\ee
separating the homogeneous background contribution, $\phi_B$, from its quantum fluctuations, $\delta \phi$, depending on the position and conformal time, $\tau= \int dt/a(t)$, with $t$ the measurable cosmic time.

The presence of fluctuations induces perturbations on the background spacetime expansion, i.e., $
g_{\mu \nu}= a^2(\tau) \left( \eta_{\mu \nu}+ h_{\mu \nu} \right)$ and $\lvert h_{\mu \nu} \rvert \ll 1$, where $a(\tau)$ is the scale factor and $\eta_{\mu \nu}$ the Minkowski metric tensor. 

We describe the slow-roll of the inflaton field in terms of a quasi-de Sitter evolution, namely \cite{Riotto:2002yw,PhysRevD.101.123522,Klaric:2022qly}
\be \label{quasids}
a(\tau)= -\frac{1}{H_I \left(\tau-2\tau_f\right)^{1+\epsilon}},
\ee
where $\tau_f$ denotes the end of the slow-roll phase. Particularly, we assume a constant Hubble parameter $H_I$, whose value is fixed at horizon crossing for the pivot scale, $k_{\rm piv}=0.05$ Mpc$^{-1}$, compatible with the Planck mission constraint \cite{Planck:2018jri}
\be \label{hubble_inf}
H_I < 2.5 \times 10^{-5} \bar{M}_{\rm pl} \simeq 6.1 \times 10^{13}\text{ GeV},
\ee
where $\bar{M}_{\rm pl}$ is the reduced Planck mass. The corresponding slow-roll parameter, $\epsilon$, can be quantified via \cite{Riotto:2002yw,Planck:2018jri},
\be \label{scapow}
\epsilon= \frac{1}{8\pi^2 P_s} \left( \frac{H_I}{\bar{M}_{\rm pl}}  \right)^2
\ee
where $P_s$ is the dimensionless scalar power spectrum, observationally constrained at $P_s=2.1 \times 10^{-9}$ for $k_{\rm piv}$. 

At linear perturbation order, a minimally coupled scalar field does not induce anisotropic stress \cite{Bassett:1998wg}, so selecting the longitudinal gauge \cite{PhysRevD.22.1882,MUKHANOV1992203} we can describe scalar perturbations in terms of a single perturbation potential\footnote{Accordingly, the perturbation metric tensor would read $h_{\mu \nu}=\text{diag}\left( 2\Phi, 2\Phi, 2\Phi, 2\Phi \right)$.} $\Phi$ 
.

The potential $\Phi$ can be derived by perturbing Einstein equations up to first order, finding \cite{Riotto:2002yw} 
\be \label{perturb}
\Phi^\prime+\mathcal{H} \Phi= \epsilon \mathcal{H}^2 \frac{\delta \phi}{\phi^\prime},
\ee
where $\mathcal{H} \equiv a^\prime/a$ and a prime denotes the derivative with respect to conformal time\footnote{In order to simplify the notation, we denote the background contribution $\phi_B$ by $\phi$ from now on.} . Rescaling inflaton fluctuations by $\delta \chi= \delta \phi a$, we can describe their dynamics via 
\be \label{flures}
\delta \chi_k^{\prime \prime}+ \left[ k^2-\frac{1}{\eta^2} \left( 2+9\epsilon- \frac{ V_{\phi \phi}}{H_I^2} \right)    \right] \delta \chi_k=0,
\ee
where $V_{\phi \phi} \equiv \partial^2 V/\partial \phi^2$ and $\eta = \tau-2\tau_f$. The general solution of Eq. \eqref{flures} reads 
\be \label{flures_sol}
\delta \chi_k(\eta)= \sqrt{-\eta} \left[ c_1(k) H_{\nu}^{(1)}(-k\eta)+c_2(k) H_{\nu}^{(2)}(-k\eta) \right],
\ee
where
 $\nu = \sqrt{\frac{9}{4}+9\epsilon-\frac{ V_{\phi \phi}}{H_I^2}}$.

Selecting now the \emph{Bunch-Davies vacuum state} \cite{Bunch:1978yq,Danielsson:2003wb,Greene:2005wk}, one finds $c_1(k)=\sqrt{\pi}e^{i(\nu +\frac{1}{2})\frac{\pi}{2}}/2$ and $c_2(k)=0$, so that the rescaled modes finally take the form
\be \label{influct}
\delta \chi_k(\eta) = \frac{\sqrt{-\pi\eta}}{2} e^{i\left( \nu+ \frac{1}{2}\right) \frac{\pi}{2}}  H^{\left(1\right)}_{\nu}\left(-k\eta\right).
\ee
On super-Hubble scales $k < aH_I$, this expression simplifies to (see e.g. \cite{Riotto:2002yw,Tsujikawa:2003jp})
\be \label{suphub}
\delta \chi_k(\eta) \simeq e^{i(\nu - \frac{1}{2})\frac{\pi}{2}} 2^{\nu-\frac{3}{2}} \frac{\Gamma(\nu)}{\Gamma(3/2)} \frac{1}{\sqrt{2k}} (-k\eta)^{\frac{1}{2}-\nu},
\ee
which will be useeful later on.

\subsection{Incorporating a fermionic spectator field} \label{sec2A}

Within the inflationary framework, we now introduce a minimally coupled Dirac field, $\psi$, whose energy density is assumed to be subdominant than $V(\phi)$ \emph{during inflation}, thus behaving as a spectator field across the entire slow-roll regime. Ensuring $\psi$ only features gravitational interactions with the background, the corresponding  Lagrangian density reads
\be \label{ferm_lag}
\mathcal{L}_S= \bar{\psi} \left( i \tilde{\gamma}^\mu \mathcal{D}_\mu -m \right) \psi,
\ee
with $m$ the field mass and $\tilde{\gamma}^\mu$ the curved spacetime Dirac matrices, with
$\left \{ \tilde{\gamma}^\mu, \tilde{\gamma}^\nu \right\}= 2 g^{\mu \nu}$.

Introducing a vierbein field in the form \cite{PhysRevD.45.4428}
\be \label{vier_inh}
\begin{aligned}
e^b_\nu= a(\tau)\left( \eta^b_\nu + h^b_\nu \right),\quad e^{a \nu}= a^{-1}(\tau) \left( \eta^{a \nu} + h^{a \nu} \right),
\end{aligned}
\ee
where latin indices are raised and lowered with the Minkowski metric, we can recover the standard flat-space Dirac matrices from
$\tilde{\gamma}^\mu= e^\mu_a \gamma^a$.

Similarly, the fermion covariant derivative $D_{\mu}$ reads \cite{Birrell_Davies_1982,Parker_Toms_2009}
$\mathcal{D}_\mu\equiv \partial_\mu+ \frac{1}{8}\left[ \gamma^a,\gamma^b \right] e^\nu_a \left( \partial_\mu e_{b \nu} -\Gamma^\lambda_{\mu \nu} e_{b\lambda}  \right)$, where $\Gamma^\lambda_{\mu \nu}$ are the standard Christoffel symbols. 

In the presence of scalar perturbations, described by Eq. \eqref{perturb}, we can expand Eq. \eqref{ferm_lag}  up to first order in $h$ by  $
\mathcal{L}_S= \mathcal{L}_S^{(0)} + \mathcal{L}_{\rm int}+ \mathcal{O}(h^2)$, 
where, by introducing the rescaled field, $\Psi=a^{3/2} \psi$, we end up with \cite{Bassett:2001jg}
\begin{subequations}
    \begin{align}
    & \mathcal{L}_S^{(0)}= \bar{\Psi} \left( i \gamma^0 \partial_\tau +i \gamma^i \partial_i -ma \right) \Psi, \label{zeroferm_lag} \\[2pt]
    & \mathcal{L}_{\rm int}= \bar{\Psi} \left( -3 \Phi \gamma^0 \partial_0-\Phi \gamma^i \partial_i - \frac{3}{2} \Phi^\prime \gamma^0 - \frac{3}{2} \gamma^i \partial_i \Phi  \right) \Psi.  \label{intferm_lag}
\end{align}
\end{subequations}
From Eqs. \eqref{zeroferm_lag}-\eqref{intferm_lag}, we can now compute the GPP by evaluating the action of fermions over the inflationary dynamics, as reported below.

\subsection{Perturbative GPP} \label{sec2B}

In the presence of small inhomogeneities, GPP can be studied within the interaction picture \cite{PhysRevD.45.4428,PhysRevD.39.389}. The fermionic field $\Psi$ can be expanded as 
\be \label{ferm_quant}
\hat{\Psi}= \int \frac{d^3k}{(2\pi)^{3/2}} e^{i {\bf k} \cdot {\bf x}} \sum_r \left[ U_r({\bf k}, \tau) b_{{\bf k},r} + V_r(-{\bf k},\tau) d^\dagger_{-{\bf k},r} \right],
\ee
where $r$ labels the spin of \emph{particles} and \emph{antiparticles}. The usual anticommutation relations plus the associated Fock space for such fermionic excitations are respectively
\begin{subequations}
\begin{align}
&\{ b_{{\bf k},r}, b^\dagger_{{\bf k^\prime},s}  \} = \{ d_{{\bf k},r}, d^\dagger_{{\bf k^\prime},s}  \}= \delta({\bf k}-{\bf k^\prime}) \delta_{rs},\label{anticomm}\\
&b_{{\bf k},r} \ket{0}= d_{{\bf k},r} \ket{0}=0,
\end{align}
\end{subequations}
where $\ket{0}$ is the Bunch-Davies vacuum state, representing a local attractor in the space of initial states for an expanding background \cite{TSBunch_1980,Brandenberger:2003vk}, whereas $\tau_i$ is the inflationary onset.

Further, the spinor mode functions $U_r$, $V_r$ satisfy the Dirac equation:
\begin{subequations}
    \begin{align}
& (i \gamma^0 \partial_\tau - {\boldsymbol \gamma} \cdot {\bf k} - ma) U_r({\bf k}, \tau) = 0,   \label{udir} \\[2pt]
& (i \gamma^0 \partial_\tau - {\boldsymbol \gamma} \cdot {\bf k} - ma) V_r(-{\bf k}, \tau) = 0   ,\label{vdir}
    \end{align}
\end{subequations}
where ${\boldsymbol  \gamma}=(\gamma^1,\gamma^2,\gamma^3)$. These equations can be solved via the ansatz 
\begin{subequations}
    \begin{align}
    & U_r({\bf k}, \tau) =  (i \gamma^0 \partial_\tau - {\boldsymbol \gamma} \cdot {\bf k} +  ma) f_k(\tau) u_r, \label{ansmod_fer1}\\
    & V_r({\bf k}, \tau) = (i \gamma^0 \partial_\tau - {\boldsymbol \gamma} \cdot {\bf k} +  ma) g_k(\tau) v_r, \label{ansmod_fer2}
    \end{align}
\end{subequations}
where $u_r= \begin{pmatrix}  \xi_r \\[3.5pt] 0    \end{pmatrix}$, $v_r= \begin{pmatrix}  0 \\[3.5pt] \xi_r    \end{pmatrix}$, and the two-component spinors $\xi_r$ are chosen to be helicity eigenstates, i.e., $
{\boldsymbol \sigma} \cdot {\bf k} = r k \xi_r$ with $ r= \pm 1$ and ${\boldsymbol \sigma}=(\sigma^1,\sigma^2,\sigma^3)$ is a vector of Pauli matrices. 

Setting $M(\tau)=ma(\tau)$, we have
\begin{subequations}
    \begin{align}
    & \left[ \frac{d^2}{d \tau^2} +k^2+M^2(\tau) -iM^\prime(\tau)  \right] f_k(\tau) = 0, \label{modefer_1}\\
    & \left[ \frac{d^2}{d \tau^2} +k^2+M^2(\tau) +iM^\prime(\tau)  \right] g_k(\tau) = 0. \label{modefer_2}
    \end{align}
\end{subequations}
Within the interaction picture, the initial vacuum state of the system evolves as $\ket{\psi}= \hat{S} \ket{0}$, 
\begin{align} \label{int_state}
\ket{\psi} = \mathcal{N} \left( \ket{0} + \frac{1}{2} \int d^3k d^3k^\prime \ket{{\bf k}, r; {\bf k}^\prime,s} \braket{{\bf k}, r; {\bf k}^\prime,s| \hat{S}^{(1)} |0} \right),
\end{align}
where $\mathcal{N} = 1+ \mathcal{O}(h^2)$ is a normalization constant and we have expanded the scattering matrix up to first order, so that
\be \label{first_s}
\hat{S}^{(1)}= i \hat{T} \int d^4x \mathcal{L}_{\rm int}.
\ee
The state in Eq. \eqref{int_state} properly defines particles in the asymptotic future $\tau \rightarrow + \infty$, i.e., when perturbations can be neglected and the adiabatic approximation holds\footnote{Such approximation is typically valid well before matter-radiation equality, see e.g. Refs. \cite{PhysRevD.101.123522,PhysRevD.101.083516}.} \cite{Birrell_Davies_1982}. In such \emph{out} region, positive-frequency (and thus physical) solutions for the field modes can be derived via Bogoliubov transformations. Similarly, we can introduce the proper ladder operators $\tilde{b}$, $\tilde{d}^\dagger$ in the \emph{out} region by
\begin{subequations}
    \begin{align}
    & \tilde{b}_{{\bf k},r} = b_{{\bf k},r} A_{k,r} - d^\dagger_{-{\bf k},r} B^*_{k,r} \label{bogo_annout} \\
    &  \tilde{d}^\dagger_{-{\bf k},r} =  d^\dagger_{-{\bf k},r} A^*_{k,r} + b_{{\bf k},r} B_{k,r},  \label{bogo_creout}
    \end{align}
\end{subequations}
where the Bogoliubov coefficients $A$ and $B$ are obtained from the matching conditions between the different epochs involved, and so they depend on how the universe expansion is modeled \cite{RevModPhys.96.045005}.
We are now ready to compute the expectation value of the \emph{out} region particle number operator,
\be \label{out_numb}
\hat{N}= \frac{1}{(2 \pi a)^{3}} \int d^3p\  \tilde{b}^{\dagger}_{{\bf p},r} \tilde{b}_{{\bf p},r},
\ee
in the state $\ket{\psi}$, being the sum of three contributions \cite{Bassett:2001jg},
\begin{align}
    & N^{(0)} = \frac{V}{(2 \pi a)^3} \int \d^3 k\  \lvert B_{k,r} \rvert^2, \label{zerodens} \\[3pt]
    & N^{(1)} = - \frac{1}{(2 \pi a)^3} \int d^3k \ {\rm Re}\left[ A_{k,r} B_{k,r} \braket{{\bf k}, r; -{\bf k},r | \hat{S}^{(1)} |0} \right], \label{firstdens}\\[3pt] 
    & N^{(2)} = \frac{1}{4(2\pi a)^3} \int d^3k d^3k^\prime\ \lvert \braket{ 0 | \hat{S}^{(1)} |{\bf k}, r; {\bf k}^\prime,s} \rvert^2 \notag \\
    &\ \ \ \ \ \ \ \ \ \ \ \ \ \ \ \ \ \ \ \ \ \ \ \ \ \ \ \ \ \ \ \   \times \left[ \lvert A_{k,r} \rvert^2 + \lvert B_{-k^\prime,s} \rvert^2  +1  \right], \label{secondens}
\end{align}
where $V$ is late time volume of the universe, i.e., as perturbations can be neglected. In particular, we notice that:
\begin{itemize}
    \item[-] The zero-order contribution of Eq. \eqref{zerodens} is the well-known particle amount arising from non-perturbative GPP due to the universe expansion \cite{PhysRev.183.1057, PhysRevD.3.346}.  
    \item[-] The first order contribution in Eq. \eqref{firstdens} emerges from the combined effect of expansion and inhomogeneities.
    \item[-] At second order, a nonzero contribution may arise even when the background expansion does not produce particles, e.g. in the limit $m \rightarrow 0$. In this case, the sole presence of inhomogeneities is responsible for conformal symmetry breaking leading to GPP, which can be quantified via the Lagrangian in Eq. \eqref{intferm_lag}.
\end{itemize}
As an important remark, a consistent treatment to perturbative GPP requires that the integrals in Eqs. \eqref{firstdens} -- \eqref{secondens} are limited to \emph{classical perturbation modes}. This can be achieved by selecting super-Hubble scales, namely $k < a H_I$, as demonstrated in Ref. \cite{Bassett:2001jg} and, furthermore, see e.g. Refs. \cite{Polarski:1995jg,PhysRevD.77.063534}.

\section{Geometric contribution to dark matter} \label{sec3}

The main drawback of non-perturbative GPP for minimally coupled Dirac fermions arises from the fact that it is strongly suppressed for small field masses \cite{Parker_Toms_2009,Chung:2011ck,Freedman_VanProeyen_2012}. 

\begin{figure}
    \centering
    \includegraphics[scale=0.48]{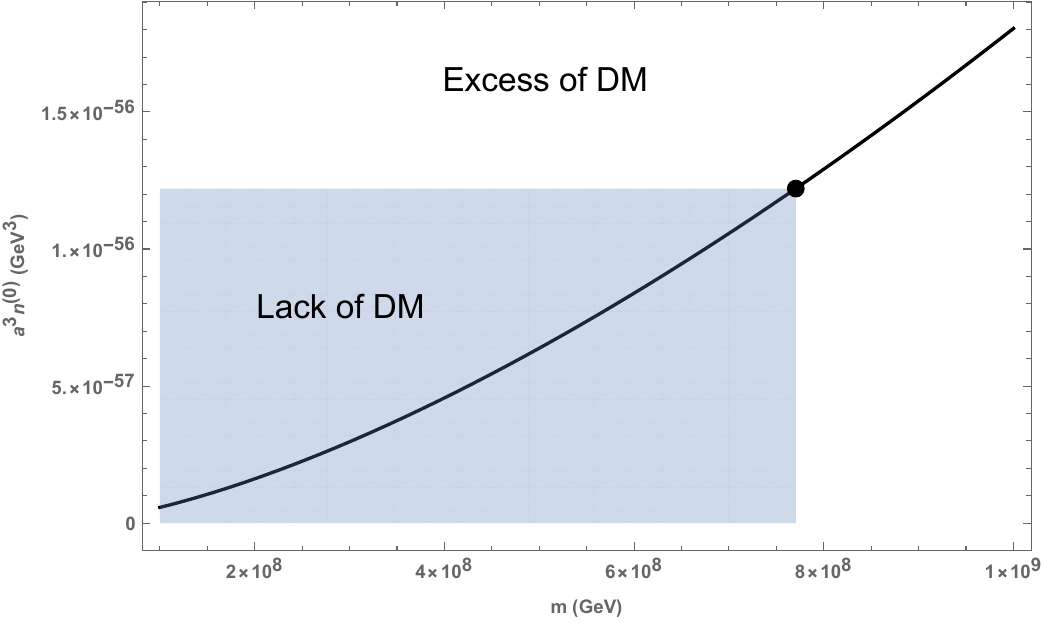}
    \caption{Comoving number density $a^3 n^{(0)}$ as function of the Dirac field mass $m$, in a model which assumes instantaneous transition from inflation to the radiation era \cite{PhysRevD.101.123522}. 
    The required DM energy density, corresponding to the black dot, is computed as $m a^3 n^{(0)}$, in agreement with a cold DM component today. Within this model, non-perturbative GPP cannot account by itself for the current DM abundance if $m \leq 7.8 \times 10^8$ GeV, which corresponds to the shaded region.}
    \label{fig_GPP}
\end{figure}

In particular, if one wants to trace back all DM to such a mechanism, the candidate fermionic particle might exhibit masses compatible with those associated with WIMPzillas  \cite{RevModPhys.96.045005}. In Fig. \ref{fig_GPP}, we show the comoving number density, $a^3 n^{(0)} \equiv a^3 N^{(0)}/V$, ensuring an \emph{instantaneous transition}\footnote{As previously mentioned, we also remark that the assumption of instantaneous transition from a de Sitter phase into a matter/radiation dominated one may lead to an overestimate of non-perturbative GPP.} from a de Sitter-like inflation, with $H_I \simeq 10^{13}$ GeV, to the radiation era \cite{PhysRevD.101.123522}. A field mass $m_{\rm GPP} \simeq 10^8$ GeV can therefore recover the measured late-time DM abundance via non-perturbative GPP.


Decreasing the window of masses is also possible within the perturbative DM production, invoking  inflationary perturbations. 

We here demonstrate that, in order \emph{to account for the entire abundance of DM, for smaller field masses, one can consider sufficiently high inflationary energy scales}. 

Indeed, recalling Eq. \eqref{quasids}, we can fix some inflationary parameters by means of Planck CMB anisotropy measurements \cite{Planck:2018jri} and, particularly:
    \begin{itemize}
        \item[-] We set the total number of inflationary e-foldings,  $N \equiv \log\left[ a(\tau_f)/a(\tau_i) \right]$, to 70, compatibly with current expectations \cite{Riotto:2002yw}.
        \item[-] We require the pivot scale lying on $k_{\rm piv}= 0.05$ Mpc$^{-1} \simeq 3.15 \cdot 10^{-40}$ GeV to leave the Hubble horizon after $N-N_{\rm piv} \equiv \log\left[ a(\tau_{\rm piv})/a(\tau_i) \right] =5$ e-foldings after the beginning of inflation. 
        \item[-] We employ the time $\tau_{\rm piv}$ from the condition $k_{\rm piv}=a(\tau_{\rm piv})H_I$, corresponding to the horizon crossing. 
    \end{itemize}
Remarkably, we emphasize the simplified scale factor,  in Eq. \eqref{quasids}, does not account for the inflationary final stages, during which the Hubble parameter is expected to drop below $H_I$.  Accordingly, a precise estimate of the total number of inflationary e-foldings for each fixed potential $V(\phi)$  would require a more refined ansatz for $a(\tau)$ when the slow-roll approximation is no longer valid.

We now focus on modes $k \lesssim a(\tau_f)H_I$, in agreement with the classicality prescription discussed above and the blue-tilted nature of the fermionic number density spectrum, which has already been proved in non-perturbative GPP scenarios\footnote{See, e.g. Ref. \cite{RevModPhys.96.045005} for a discussion on this point.}. Consequently, the contribution of the modes $k \ll a(\tau_f)H_I$ can typically be neglected.

In the limit $m < m_{GPP} \ll H_I$,  the mass terms in Eqs. \eqref{modefer_1}-\eqref{modefer_2} become negligible for modes leaving the horizon during the slow-roll latest stages. There,  $N^{(0)},N^{(1)} \simeq 0$ and the second-order term of Eq. \eqref{secondens} acquires the simplified form \cite{PhysRevD.45.4428, Bassett:2001jg}:
\be \label{secord_weyl}
N^{(2)}= \frac{1}{160 \pi a^3} \int \frac{d^4k}{(2\pi)^4} \theta(k^0) \theta(k^2) \lvert \tilde{C}^{abcd}(k) \rvert^2,
\ee
where $k_0 = \left( k^2 + M^2 \right)^{1/2}$, moreover introducing the Fourier transform 
$\tilde{C}^{abcd}(k)= \int d^4x \ e^{i k \cdot x} C^{abcd}(x)$ of the Weyl tensor, $C^{abcd}$.

For scalar inflaton perturbations, Eq. \eqref{secord_weyl} provides the number density,
\begin{align} \label{numbdens2}
n^{(2)}\equiv \frac{N^{(2)}}{V}= & \frac{1}{15 (2\pi)^4a^3} \int_{k_0^{IR}}^{a(\tau_*)H_I} dk_0 \int_{k^{IR}}^{k_0} dk k^6 \notag \\[2pt]
& \times \left \lvert \int_{\tau_*}^{\tau_f} d\tau e^{i k_0 \tau} \Phi_k(\tau) \right \rvert^2,
\end{align}
where the perturbation modes $\Phi_k$ can be derived from Eq. \eqref{perturb} once the inflationary potential has been selected. 

Here, we assume that the possible additional contribution due to preheating is  neglected\footnote{Perturbative GPP from preheating metric perturbations has been studied in Ref. \cite{Bassett:2001jg}, showing that the final amount of particles is heavily dependent on the nature of the inflaton interactions with other quantum fields after inflation.}, thus obtaining a lower bound for the total particle number density.

Furthermore, we highlight that the time, $t_*$, might be selected so that all modes of interest are  super-Hubble within the time interval $[\tau_*,\tau_{f}]$, to ensure the classicality of $\Phi_k$. Due to the blue-tilted nature of the number density spectrum, we can safely assume $k_0^{IR} \simeq k^{IR} \gg a(\tau_i)H_I$, thus focusing only on modes that leave the Hubble horizon at $\tau \lesssim \tau_*$.

\begin{table*}[ht!] 
\centering 
\resizebox{13cm}{!}{\begin{tabular}{|c|cc|cc|} 
\hline\hline
\multirow{2}{*}{$H_I$ ($10^{13}$ GeV)} & \multicolumn{2}{c|}{ Starobinsky} & \multicolumn{2}{c|}{ Quadratic hilltop} \\ \cline{2-5} 
 & \multicolumn{1}{c|}{\ $a^3n^{(2)}$ ($10^{-52}$ GeV$^{3}$)\ } & \ m ($10^6$ GeV) \  & \multicolumn{1}{c|}{\ $a^3n^{(2)}$ ($10^{-52}$ GeV$^{3}$)\ } & \ m ($10^6$ GeV) \  \\ \hline\hline
 4.0 & \multicolumn{1}{c|}{$0.042$} & $2.230 $ & \multicolumn{1}{c|}{$0.00194$} & $48.4$ \\ \hline
4.5 & \multicolumn{1}{c|}{$0.112$} & $0.838$ & \multicolumn{1}{c|}{$0.00328$} & $28.7$ \\ \hline
5.0 & \multicolumn{1}{c|}{$0.233$} & $0.404$ & \multicolumn{1}{c|}{$0.00531$} & $17.7$ \\ \hline
5.5 & \multicolumn{1}{c|}{$0.550$} & $0.171$ & \multicolumn{1}{c|}{$0.00836$} & $11.2$ \\ \hline
6.0 & \multicolumn{1}{c|}{$1.030 $} & $0.091$ & \multicolumn{1}{c|}{$0.01290$} & $7.3 $ \\ \hline\hline
\end{tabular}}
\caption{\enquote{Geometric} comoving number density and corresponding mass for a fermionic DM candidate, playing the role of a spectator field during inflation. Starobinsky inflation and a quadratic hilltop potential have been selected to compute the number of particles produced in the interval $[\tau_1,\tau_f]$, for super-Hubble modes $k_1/100 \leq k \leq k_1$, with $k_1=a(\tau_1)H_I$. }
\label{tab_DM}
\end{table*}

In the following, we refine our calculations for key inflationary scenarios, examining some among the most promising large and small-field models.

\begin{itemize}
\item[-] \emph{Large field Starobinsky potential.} The Starobinsky potential, characterized by the inclusion of the quadratic term $R^2$ in the Einstein-Hilbert action, currently represent the leading candidate to describe the inflationary phase. The corresponding inflationary potential can be expressed in the form \cite{Planck:2018jri}
\be \label{staropot}
V(\phi)= \Lambda^4\left( 1-e^{-\sqrt{2/3} \phi/\bar{M}_{\rm pl}} \right)^2,
\ee
where $\Lambda \simeq 10^{16}$ GeV quantifies the energy scales of inflation. Recalling Eq. \eqref{infans}, the slow-roll background dynamics of $\phi$ is described by
\be \label{}
3 \mathcal{H} \phi^\prime \simeq - V_{,\phi} a^2,
\ee
where $a$ has been defined in Eq. \eqref{quasids}. Once the inflaton background evolution and the corresponding fluctuation dynamics have been derived, metric perturbations are obtained from Eq. \eqref{perturb}. On super-Hubble scales, the perturbation potential simplifies to
\be \label{perturb_suph}
\Phi \simeq \epsilon \mathcal{H} \frac{\delta \phi}{\phi^\prime},
\ee
which readily allows to compute Eq. \eqref{numbdens2} via the fluctuation modes of Eq. \eqref{suphub}. It can be shown that
\be \label{avg_fluc}
\phi(\tau_i) \gg \langle \delta \phi^2 \rangle^{1/2} \simeq \frac{H_I}{2\pi},
\ee
thus preserving our perturbative approach throughout the slow-roll phase.

In Table I, we show the comoving number density $a^3n^{(2)}$ at the end of the inflationary phase, for fixed values of the Hubble parameter $H_I$. We set $\tau_*=\tau_1$, where $\tau_1$ corresponds to the last inflationary e-folding, namely $\ln[a(\tau_f)/a(\tau_1)]=1$.

Accordingly, in our computation we do not include the contribution of those modes which leave the Hubble horizon after $\tau_1$. 
This implies that our results provide a lower bound on the total amount of particles produced via perturbative GPP in slow-roll\footnote{Back-reaction effects may in principle reduce the total particle amount, despite their contribution during slow-roll should be negligible, at least from a classical perspective. See e.g. Ref. \cite{PhysRevD.107.103512}.}. 

Once GPP has completed, if no additional particle-number-changing reactions are active for the field $\psi$, the comoving number density is conserved, thus implying $a^3 n^{(2)}= a_0^3 n_0^{(2)}$, where $a_0$ gives the scale factor today and $n_0$ the corresponding second-order number density.  Setting, as usual, $a_0 \simeq 1$, and denoting by $\tau_0$ the conformal time today, we observe that
\be \label{com_freq}
k_0(\tau_0)= \left( k^2+M^2(\tau_0) \right)^{1/2} \simeq m 
\ee
for all modes leaving the Hubble horizon during inflation, provided $m \geq 1$ eV.  Accordingly, the spectator field $\psi$ is a nonrelativistic (cold) species today, for sufficiently large field masses. Its current energy density simplifies to $\rho_\psi \simeq mn_0^{(2)}$ and we can estimate the corresponding cosmological energy fraction  by evaluating the ratio $\Omega_\psi\equiv \rho_\psi/\rho_c$, where $\rho_c=3 \bar{M}^2_{\rm pl}H^2_0$ is the current critical density of the universe, with $H_0=2.1 h \times 10^{-42}$ GeV and $h \simeq 0.674$ \cite{Planck:2018jri}. 

In Tab. \ref{tab_DM}, we display the expected fermionic field mass if all the DM present in the universe is traced back to perturbative GPP for the spectator field $\psi$, namely we set $\Omega_\psi \equiv \Omega_{\rm DM} \simeq 0.12/h^2$, corresponding to $\rho_\psi \simeq 9.40 \times 10^{-48}$ GeV$^{4}$.

\item[-] \emph{Small field quadratic hilltop potential.} As an alternative scenario, we discuss the small-field quadratic hilltop inflation, with corresponding potential
\be \label{hill_pot}
V(\phi)= \Lambda^4\left[1- \left( \frac{\phi}{\mu_2} \right)^2  \right],
\ee
where $\mu_2$ is constrained by Planck observations and we select the same energy scales of the Starobinsky scenario. Following the same steps discussed above, we can compute the number density $n_0$ today and the corresponding mass of the DM candidate, which are both summarized in Tab. \ref{tab_DM} for $\mu_2= 5\  M_{\rm pl}$. We observe that the amount of produced particles is similar with respect to the large-field Starobinsky scenario, and it can be shown that larger densities are obtained for larger $\mu_2$.

\end{itemize}

Accordingly, we have shown that, within both scenarios, a fermionic spectator field with mass $10^5 \lesssim m \lesssim 10^7$ GeV may represent a plausible cold DM candidate, exhibiting the correct energy density today, in agreement with Planck data\footnote{We highlight that the expected DM energy density, namely $\rho_c \Omega_{\rm DM}$, is independent of the exact $h$ value.}.

\subsection{Including particle contribution during reheating} \label{sec3A}

As shown by Eq. \eqref{numbdens2}, we have restricted our computation of particle densities to the slow-roll phase of inflation. However, an additional contribution to $n^{(2)}$ would inevitably arise from the dynamics of the perturbation potential modes $\Phi_k$ at the end of inflation and during reheating, when the inflaton field is expected to coherently oscillate around the minimum of its potential and its energy is gradually transferred to other particles, which then scatter and thermalize to form the primordial plasma \cite{Kofman:1997yn, Bassett:2005xm,Allahverdi:2010xz}.

The oscillating nature of the inflaton field during reheating suggests  a negligible contribution to the number density in Eq. \eqref{numbdens2}, independently from the duration of the reheating stage. Furthermore, field modes that leave the Hubble horizon during the latest stages of slow-roll are expected to be again in sub-Hubble form soon after the end of inflation, thus violating the above-discussed classicality condition for perturbation modes.

A non-negligible contribution to $n^{(2)}$ may arise from an initial phase of preheating, during which metric perturbations undergo parametric amplification. Both preheating and reheating are highly model-dependent, so this additional contribution depends on the eventual coupling(s) between the inflaton field and Standard Model fields \cite{Bassett:2001jg}. If we neglect such couplings, the dominant mechanism for perturbative GPP during reheating may emerge from inflaton annihilation by s-channel graviton exchange\footnote{Alternatively, DM might be produced in reheating via gravity-mediated thermal freeze-in, see e.g. Refs. \cite{PhysRevLett.116.101302,Tang:2016vch,Tang:2017hvq,PhysRevD.99.063508}. However, this mechanism is expected to be subdominant with respect to inflaton annihilation, since more energy is available in the inflaton field at the end of inflation \cite{PhysRevD.105.075005}.} \cite{Ema:2015dka, PhysRevD.94.063517,Tang:2017hvq,PhysRevD.103.115009, Mambrini:2022uol}. Such gravitational portal \cite{PhysRevD.105.075005} may significantly contribute to the total dark matter abundance, provided the mass of the fermionic DM candidate and the reheating temperature are both sufficiently large. Accordingly, this additional process should be taken into account in order to obtain a more refined estimate of the DM candidate mass.

\section{Physical interpretation and final remarks}\label{sec4}

Our findings suggested that a Dirac spectator field, in inflationary stages, may represent a viable DM candidate to generate the entire abundance of DM for field masses down to $m \simeq 10^5$ GeV, commonly excluded when inflationary perturbations have not been considered in GPP processes. 

In so doing, we opened a new window of allowed mass candidates, enabling, at the same time, DM to be fully produced by the mechanism of perturbative GPP from inhomogeneities.

In particular, the non-perturbative contribution arising from background expansion, according to which the homogeneous gravitational field transfers part of its energy to the spectator field, has been shown to be typically negligible for a Dirac field with $m \leq 10^8$ GeV, assuming inflationary energy scales $H_I \simeq 10^{13}$ GeV. Conversely, the gravitational coupling between inflationary perturbations and the spectator field may lead to significant particle production even for smaller field masses and, summarizing: 
\begin{itemize}
    \item[-] Perturbative GPP from inhomogeneities may provide a non-negligible contribution to the number density of particles produced via \emph{purely gravitational effects}.
    \item[-] When dealing with minimally coupled fermionic fields, with mass $m \leq 10^8$ GeV, non-perturbative GPP \emph{cannot account by itself for producing viable DM candidates}, even assuming  instantaneous transition between inflation and the radiation-dominated era.
    \item[-] If the inflationary energy scales, which we fix at the horizon crossing of the pivot scale $k_{\rm piv}=0.05$ Mpc$^{-1}$, are sufficiently high, namely $H_I \geq 4 \times 10^{13}$ GeV, \emph{perturbative GPP can produce the required DM abundance during slow-roll for masses $10^5 \lesssim m \lesssim 10^7$ GeV, for which the non-perturbative mechanism is inefficient}. 
    \item[-] Perturbative GPP has been worked out for large and small-field inflationary models. It appeared slightly larger in the case of Starobinsky inflation than hilltop potential, despite this depends on $\mu_2$ values.
    \item[-] A more refined description of the latest inflationary stages (i.e., when the slow-roll quasi-de Sitter approximation is no longer valid) may provide a more precise estimate of the particle number density, and thus of the DM candidate mass.
\end{itemize}

Moreover, to properly compute the particle amount, we focused on super-Hubble modes, exploiting their classical features, and for masses $m \ll H_I$ we obtained the particle number density via the Weyl tensor associated with the inflationary metric perturbations, arising from the scalar inflaton fluctuations.

Accordingly, the spectrum of such \enquote{geometric} particles looked strongly blue-tilted, namely the modes with larger contribution are those leaving the Hubble horizon during the latest slow-roll stages.

Last but not least, additional gravitational processes during reheating, e.g. amplification of metric perturbations, inflaton annihilation or gravity-mediated thermal freeze-in, may enhance the total number density of produced particles, despite these additional contributions strongly depend on the reheating parameters.

In future efforts, we primarily plan to further investigate the nature of particles arising from perturbative GPP processes. We also aim to refine our DM mass estimate by further studying possible contributions arising from the latest stages of inflation and reheating, also highlighting their dependence on the reheating parameters. Furthermore, perturbative GPP processes from vacuum may provide relevant number densities also in case of spectator fields with different fundamental properties such as spin, cross-section and so on. Additional exotic fields, non-minimal couplings and mixtures of more complicated sectors will be also objects of future investigations.

\section*{Acknowledgments}
OL expresses his gratitude to Thibault Damour for insightful discussions on the topic of geometric particle production. The authors are also in debit with Stefano Mancini, Ugo Moschella and Roland Triay for interesting debates on the topic of field theories applied to particle production in the primordial universe. OL acknowledges support by the  Fondazione  ICSC, Spoke 3 Astrophysics and Cosmos Observations. National Recovery and Resilience Plan (Piano Nazionale di Ripresa e Resilienza, PNRR) Project ID $CN00000013$ ``Italian Research Center on  High-Performance Computing, Big Data and Quantum Computing" funded by MUR Missione 4 Componente 2 Investimento 1.4: Potenziamento strutture di ricerca e creazione di ``campioni nazionali di R\&S (M4C2-19)" - Next Generation EU (NGEU).

\bibliographystyle{unsrt}
\bibliography{biblio}

\end{document}